\documentstyle[aps,12pt]{revtex}

\begin{document}
\title{Comment on ''Alternating spin and orbital dimerization and spin-gap
formation in coupled spin-orbital systems'' }
\author{R. J.Radwanski}
\address{C{enter for Solid State Physics, S}$^{{nt}}${\ Filip 5, 31-150 Krak%
\'{o}w,}\\
{Inst. of Physics, Pedagogical University, 30-084 Krak\'{o}w, POLAND. }}
\author{Z.Ropka}
\address{{Center for Solid State Physics, S}$^{{nt}}${\ Filip 5, 31-150 Krak%
\'{o}w,}\\
POLAND.}
\maketitle

\begin{abstract}
\end{abstract}

\pacs{}

In Ref. 1 Pati et al. have discussed a novel mechanism for obtaining a
spin-gap state through the creation of a coherent spin-orbital structure in
systems with orbital degeneracy. The authors consider the effective
spin-pseudo-spin Hamiltonian. Their results authors have applied to NaV$_2$O$%
_5$ and Na$_2$Ti$_2$Sb$_2$O. In both of these systems an ion with one 3d
electron exists (V$^{4+}$ and Ti$^{3+}$ ions are 3d$^1$ systems).

By this Comment we would like to point out [2] that, though such
consideration can be quite sophisticated, the consideration of a
spin-pseudo-spin Hamiltonian has nothing in common with the reality. It
means that such the consideration are useless as far as the properties of
real systems are concerned.

It is a text-book knowledge [3,4] that 1 3d electron has to be described by
the spin quantum number S=1/2 and the orbital quantum number L=2. Moreover,
these two quantum numbers are coupled via the spin-orbit coupling, that,
though weak, surely exists.

According to us it is the spin-orbit coupling that ties together the spin
and orbital degrees of freedom [5] in contrary to a fully artificial new
mechanism proposed in Ref. 1. The spin-orbit coupling is well founded in
physics in contrary to this new mechanism. Following the Occam's razor
principle the well founded mechanism is superior.

The authors write ''Thus the ground state of Ti contains one d electron in a
doubly degenerate orbital'' (p. 5409, left column, 4 line top). According to
us there is no chance for the physical realization of the ground state of Ti
with one d electron in a doubly degenerate orbital. In reality it will
appear a Jahn-Teller effect that removes this orbital degeneracy. Moreover,
a quite simple calculations prove that the tetragonal distortion discussed
by authors on p. 5408, right column for Na$_{2}$Ti$_{2}$Sb$_{2}$O, in the
presence of the always existing intra-atomic spin-orbit coupling, produces
already the orbital singlet (see e.g. Ref. 6) in contrary to the speculation
of Ref. 1 about the doublet e$_{g}$.

Thus the condition required for the authors' model cannot be physically
realized.

In conclusion, the model discussed by authors is completely artificial as
the condition about the doublet orbital ground state required for the
authors' model cannot be physically realized. Moreover, the theoretical
calculations of Ref. 1 are fully invalidated by neglection of the spin-orbit
coupling and the employing of the artificial quantum numbers.

\end{document}